\documentclass[pre,apstwocolumn,showpacs,floatfix,nofootinbib]{revtex4}
\usepackage{graphics}
\usepackage{epsfig}
\usepackage{amsmath}

\begin{document}

\title{Extremely large scale simulation of a Kardar-Parisi-Zhang 
model using graphics cards}

\author{Jeffrey Kelling (1) and G\'eza \'Odor (2)}

\affiliation{(1) Institute of Ion Beam Physics and Materials Research \\
Helmholtz-Zentrum Dresden-Rossendorf \\
P.O.Box 51 01 19, 01314 Dresden, Germany \\
(2) Research Institute for Technical Physics and Materials Science, \\
P.O.Box 49, H-1525 Budapest, Hungary}

\begin{abstract}
The octahedron model introduced recently has been implemented onto 
graphics cards, which permits extremely large scale simulations
via binary lattice gases and bit coded algorithms. 
We confirm scaling behavior belonging to the two-dimensional 
Kardar-Parisi-Zhang universality class and find a surface 
growth exponent: $\beta=0.2415(15)$ on $2^{17}\times 2^{17}$ 
systems, ruling out $\beta=1/4$ suggested by field theory. 
The maximum speedup with respect to a single CPU is 240. 
The steady state has been analyzed by finite size scaling and a growth
exponent $\alpha=0.393(4)$ is found.
Correction-to-scaling exponents are computed and the power-spectrum
density of the steady state is determined. We calculate the universal 
scaling functions, cumulants and show that the limit distribution 
can be obtained by the sizes considered. We provide numerical fitting 
for the small and large tail behavior of the steady state 
scaling function of the interface width.
\end{abstract}
\maketitle

%%%%%%%%%%%%%%%%%%%%%%%%%%%%%%%%%%%%%%%%%%%%%%%%%%%%%%%%%%%%%%%%%%%%%%%%%%
\section{Introduction}
%%%%%%%%%%%%%%%%%%%%%%%%%%%%%%%%%%%%%%%%%%%%%%%%%%%%%%%%%%%%%%%%%%%%%%%%%%

The research of the nonlinear stochastic differential equation and
universality class introduced by Kardar, Parisi and Zhang (KPZ) \cite{KPZeq} 
is in the forefront of interest nowadays again \cite{KPZpede,CorwinUof}.
This is largely due to the progress in exact solutions for various 
one-dimensional realizations and initial conditions 
(see for example \cite{Joh00,PraSp00,Fer05,Sasa04,CD11}).
This equation can describe the dynamics of simple growth processes in the 
thermodynamic limit \cite{barabasi,H90}, randomly stirred fluid 
\cite{forster77}, directed polymers in random media
\cite{kardar87} dissipative transport \cite{beijeren85,janssen86},
and the magnetic flux lines in superconductors \cite{hwa92}. 
Due to the mapping onto the Asymmetric Exclusion Process (ASEP) 
\cite{Rost81} in one dimension it is also a fundamental model of 
non-equilibrium particle system \cite{Obook08}. 
The KPZ equation specifies the evolution of the height function 
$h(\mathbf{x},t)$ in the $d$ dimensional space
\begin{equation}  \label{KPZ-e}
\partial_t h(\mathbf{x},t) = v + \sigma_2\nabla^2 h(\mathbf{x},t) +
\lambda(\nabla h(\mathbf{x},t))^2 + \eta(\mathbf{x},t) \ .
\end{equation}
Here $v$ and $\lambda$ are the amplitudes of the mean and local growth
velocity, $\sigma_2$ is a smoothing surface tension coefficient and $\eta$
roughens the surface by a zero-average, Gaussian noise field exhibiting
the variance
$\langle\eta(\mathbf{x},t)\eta(\mathbf{x^{\prime}},t^{\prime})\rangle = 2 D
\delta^d (\mathbf{x-x^{\prime}})(t-t^{\prime})$.
The letter $D$ denotes the noise amplitude and $\langle\rangle$
means the distribution average.
The equation is solvable in $\left( 1+1\right) d$ due to the Galilean 
symmetry \footnote{The invariance of Eq. (\ref{KPZ-e}) under an infinitesimal 
tilting of the interface, resulting in $\alpha + z =2$}, 
\cite{forster77} and an incidental fluctuation-dissipation symmetry 
\cite{kardar87}, while in higher 
dimensions approximations are available only. The model exhibits diverging
correlation length, thus scale a invariance, that can be understood by 
the steady current in the ASEP model corresponding to the up-down 
anisotropy of KPZ. Therefore KPZ equation has been investigated by 
renormalization techniques \cite{SE92,FT94,L95}.
As the result of the competition of roughening and smoothing terms,
models described by the KPZ equation exhibit a roughening phase transition
between a weak-coupling regime ($\lambda <\lambda _{c}$) and a strong
coupling phase. The strong coupling fixed point is inaccessible by
perturbative renormalization group (RG) method.
Therefore, the KPZ phase space has been the subject of controversies
for a long time \cite{MPP,MPPR02,F05} and the strong coupling fixed 
point has been located by non-perturbative RG very recently \cite{CDDW09}.

Discretized versions of KPZ have also been studied a lot 
(\cite{FT90,GCT09,TNM10}, for a review see \cite{barabasi}).
Recently we have shown \cite{asep2dcikk,asepddcikk} the mapping between
the KPZ surface and the ASEP \cite{kpz-asepmap,meakin} can 
straightforwardly be extended to higher dimensions.
In 2+1 dimensions the mapping is just the simple extension of the
rooftop model to the octahedron model as can be seen on Fig. 2 of
\cite{asep2dcikk}.
The surface built up from the octahedra can be described by the
edges meeting in the up/down middle vertexes. The up edges in the
$x$ or $y$ directions are represented by $\sigma_{x/y} = +1$-s, 
while the down ones by $\sigma_{x/y}-1$ in the model. 
This can also be understood as a special $2d$ cellular automaton 
\cite{Wolfram} with the generalized Kawasaki updating rules
\begin{equation}\label{rule}
\left(
\begin{array}{cc}
   -1 & 1 \\
   -1 & 1 
\end{array}
\right)
 \overset{p}{\underset{q}{\rightleftharpoons }}
\left(
\begin{array}{cc}
   1 & -1 \\
   1 & -1 
\end{array}
\right)
\end{equation}
with probability $p$ for attachment and probability $q$ for detachment.
We have confirmed that this mapping, using the parametrization: 
$\lambda = 2 p/(p+q)-1$, reproduces the one-point functions of the 
continuum model. 
This kind of generalization of the ASEP model can be regarded
as the simplest candidate for studying KPZ in $d>1$: a 
one-dimensional model of self-reconstructing $d$-mers 
\cite{BGS07} diffusing in the $d$-dimensional space.
Furthermore this lattice gas can be studied by very efficient 
simulation methods. 

Now we implement dynamic, bit-coded simulations of the conserved lattice 
gas models for graphics cards (GPUs), allowing very large system 
sizes $L\times L$. 
The surface heights are reconstructed from the slopes
\begin{equation}
h_{i,j} = \sum_{l=1}^i \sigma_x(l,1) + \sum_{k=1}^j \sigma_y(i,k)
\end{equation}
and the squared interface width
\begin{equation} 
\label{Wdef}
W^2(L,t) = \frac{1}{L^2} \, \sum_{i,j}^L \,h^2_{i,j}(t)  -
\Bigl(\frac{1}{L} \, \sum_{i,j}^L \,h_{i,j}(t) \Bigr)^2 \ .
\end{equation}
was calculated at certain sampling times ($t$).
In the absence of any characteristic length, growth processes are expected to
follow power-law behavior and the surface can be described by
the {\em Family-Vicsek}~\cite{family} scaling law:
\begin{equation}
\label{FV-forf}
W(L,t) \simeq L^{\alpha} f(t / L^z),
\end{equation}
with the universal scaling function $f(u)$
\begin{equation}
\label{FV-fu}
f(u) \sim
\left\{ \begin{array}{lcl}
     u^{\beta}     & {\rm if} & u \ll 1 \\
     {\rm const.} & {\rm if} & u \gg 1
\end{array}
\right .
\end{equation}
Here $\alpha$ is the roughness exponent of the stationary regime,
when the correlation length has exceeded the system size $L$; and
$\beta$ is the growth exponent, describing the intermediate time
behavior. The dynamical exponent $z$ is just the ratio
\begin{equation}\label{zlaw}
z = \alpha/\beta \ .
\end{equation}

%%%%%%%%%%%%%%%%%%%%%%%%%%%%%%%%%%%%%%%%%%%%%%%%%%%%%%%%%%%%%%%%%%%%%%%%%%
\section{Bit-coded GPU simulations}
%%%%%%%%%%%%%%%%%%%%%%%%%%%%%%%%%%%%%%%%%%%%%%%%%%%%%%%%%%%%%%%%%%%%%%%%%%

The height of each surface site is thoroughly determined by two slopes, along
the $x$ and $y$ axes respectively, whose absolute values are restricted to
unity. Thus at each site two bits of information are required, hence a chunk of
$4 \times 4$ sites is encoded in one 32-bit word.

\begin{figure}[h!t]
\begin{center}
\epsfxsize=70mm
\epsffile{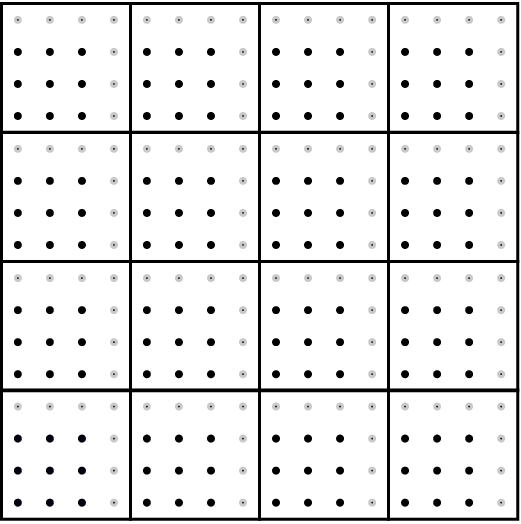}
\caption{Sketch of the dead border domain decomposition scheme employed. Lattice
sites are indicated by dots, where the grey dots represent inactive sites. Since
only two slopes relating any sites to its nearest neighbors are stored off-site
only two edges are inactive.}
\label{fig:domaindecomposition}
\end{center}
\end{figure}

Two slightly different layers of parallelization are used that reflect the two
layered compute architecture provided by GPUs~\cite{NVCPG}: not communicating
blocks at \emph{device level} and communicating threads at \emph{block level}.
Both layers use domain decomposition with dead borders, i.e. conflicts at the
subsystem borders are avoided simply by not updating them (see Fig.~\ref
{fig:domaindecomposition}). A random translation is applied to the origin of
the decomposition periodically to preserve statistics. To avoid having to deal
with non 32-bit aligned memory these translations are restricted to multiples
of four sites. 

The complete system is stored in global device memory, each block cell is copied
into the block-local shared memory for precessing. Thus moving the origin of the
device level decomposition results in cutting out the proper piece of the
system taking the periodic boundary conditions into account. Moving the origin
at each Monte Carlo step (MCS), i.e. by one overall update of the system, 
proved to be sufficient and the results are undistorted.

The size of a thread cell is set to be $8\times8$ sites, the smallest possible
choice, to maximize the number of threads per block. Due to this small subsystem
size a new origin for the block level decomposition is picked at each update
attempt, thus 64 times per MCS. Additionally the borders are not dead but
delayed, i.e. when a thread picks the border of its cell to change it waits for
the threads updating bulk sites to finish their updates. Corner sites are
further delayed.

If a thread hits a site belonging to a block border it does nothing, slightly
reducing the actual number of update attempts per MCS. This is a minor effect,
the ratio of block border to overall system size is approximately $1/128$,
and impacts the pre-factors of the scaling, but not exponents.

For random number generation each thread uses a 32-bit, linear congruential
random number generator (LCRNG) with different seed~\cite{newman99a}. Similar
generators were previously successfully applied in GPU simulations of the Ising
model~\cite{preis09}. Depending on the system size the generators have to be
periodically reseeded, which potentially introduces the same correlations 
as using multiple generators in parallel. 
However, since no deviations from the earlier CPU results have been 
observed, we assume this to not disturb the statistics. Correlations
resulting from parallel usage could only have local effects on the 
updates of a block cell. 
Moving the origin of the block level decomposition should effectively 
destroy such correlations. By the same argument reseeding the generators, 
using a Mersenne Twister generator \cite{MT} running on the CPU, has 
no negative effect at all. 
Part of the results were double-checked, employing a skip-ahead 64-bit
LCRNG~\cite{2011arXiv1101.1427W} with no need for reseeding.

\begin{figure}[h!t]
\begin{center}
\epsfxsize=70mm
\epsffile{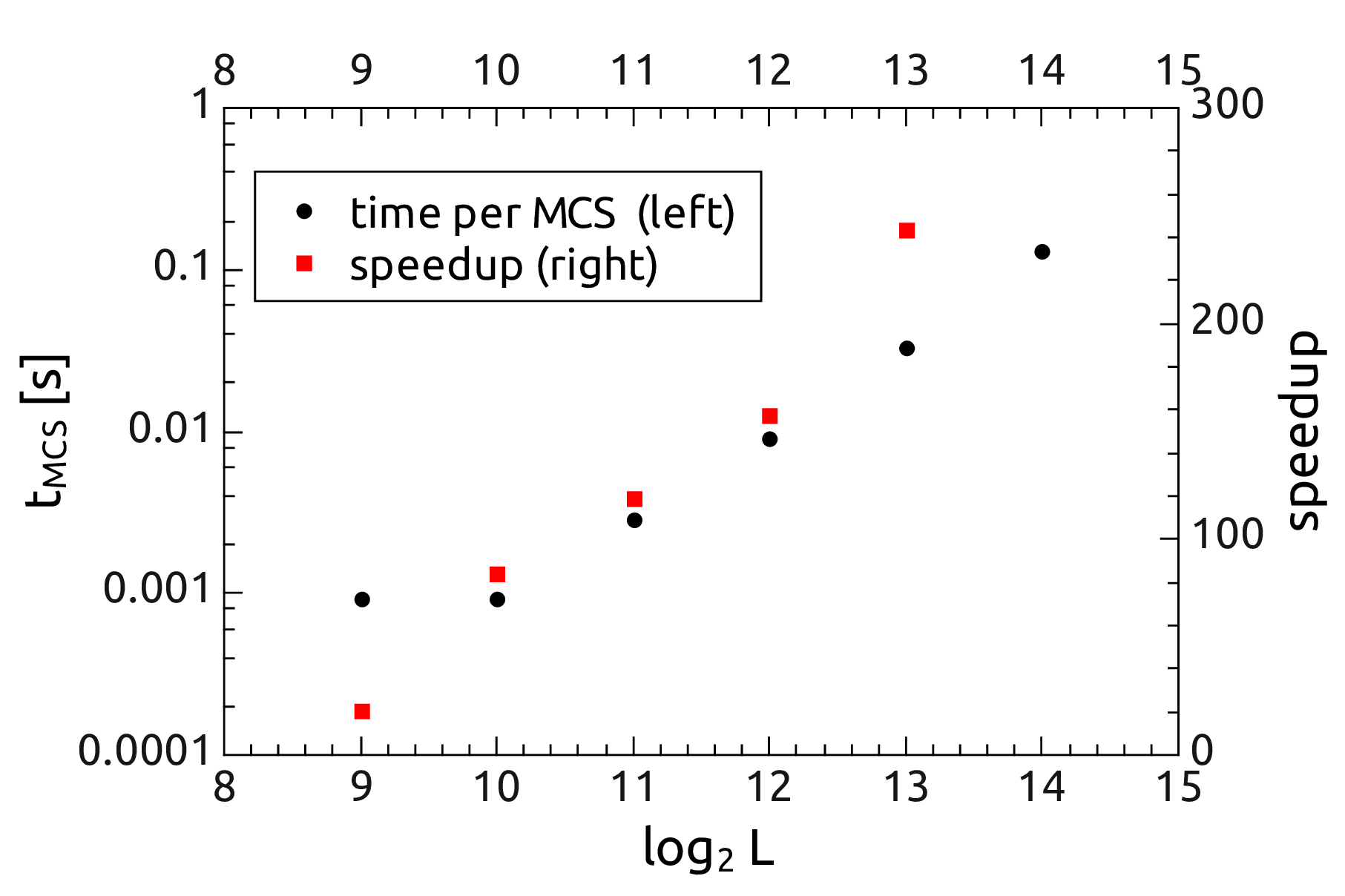}
\caption{(Color online)
Time in seconds for one MCS on a C2070 (black bullets, left axis).
Speedup over a non parallel implementation on a Core i5 750@3GHz (red squares,
right axis). The jump in speedup between $\log_2 L = 12$ and $13$ results from
the system exceeding the CPUs L3 cache at this point. The maximum speedup factor
achieved was roughly 240.}
\label{fig:benchmark}
\end{center}
\end{figure}

The simulations were performed on a C2070 GPU with 6GB graphics memory, which
allowed for a maximum system size: $2^{17}\times 2^{17}$ (4GB of memory
required). 
Figure~\ref{fig:benchmark} shows benchmark results for the simulation.

The benchmarks only consider bare simulation times and exclude the
time needed to transfer data between host and device. For large sizes,
where the system exceeds the CPUs L3 cache, the performance drops
significantly, this could be avoided in a CPU implementation using
domain decomposition designed to optimize memory access of the CPU cache.

%%%%%%%%%%%%%%%%%%%%%%%%%%%%%%%%%%%%%%%%%%%%%%%%%%%%%%%%%%%%%%%%%%%%%%%%%%
\section{Surface growth scaling}
%%%%%%%%%%%%%%%%%%%%%%%%%%%%%%%%%%%%%%%%%%%%%%%%%%%%%%%%%%%%%%%%%%%%%%%%%%

We have run $10-1000$ independent simulations for sizes: 
$L=2^8, 2^9, 2^{10}, 2^{11}, 2^{12}, 2^{13}, 2^{16}, 2^{17}$
by starting from half filled (striped) lattice gases.
This causes an intrinsic width of the initial zig-zag surface state, 
with $W^2(L,0)=1/4$, a leading order constant correction to scaling, 
that we subtract at the beginning of the scaling analysis. 
The time between measurements increases exponentially
\begin{equation}
t_{i+1} = (t_i  + 10) \cdot \mathrm{e}^m, \quad \text{with} \quad m > 0, \quad t_0 = 0,
\end{equation}
where the program calculates and writes out the width of the 
surface. We used $m=0.01$ to study the growth in
larger systems, and $m=0.001$ or $m=0.0001$ to collect more 
data points in the steady state. 
By the scaling analysis we disregarded the initial time region: 
$t < t_\mathrm{mic} \simeq 50$, when basically an uncorrelated, random
deposition process goes on.
The growth is expected to follow simple scaling~(\ref{FV-fu})
asymptotically and we assume corrections in the form
\begin{equation}\label{betacorr}
W(t,L\to\infty) = b t^{\beta}(1 + b_0 t^{\phi_0} + b_1 t^{\phi_1} ...) \ .
\end{equation}
For finite system, when the correlation length exceeds $L$, 
the growth crosses over to a saturation, with the expected scaling
behavior 
\begin{equation}\label{alphacorr}
W(t\to\infty,L) = a L^{\alpha}(1 + a_0 L^{\omega_0} + a_1 L^{\omega_1} ...) \ .
\end{equation}

\begin{figure}
\begin{center}
\epsfxsize=70mm
\epsffile{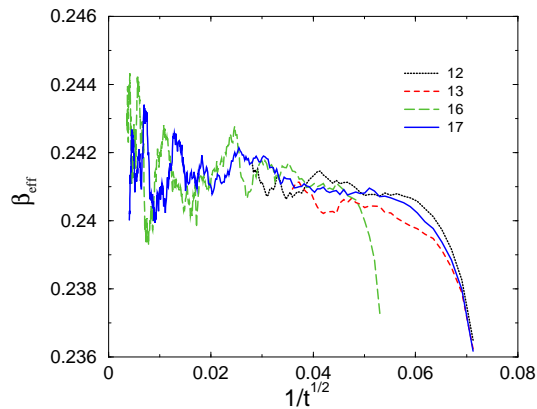}
\caption{(Color online)
Local slopes of the surface growth for different sizes
($L=2^{12}, 2^{13}, 2^{16}, 2^{17}$). Averaging was done over 20-100
independent runs.}
\label{slopes}
\end{center}
\end{figure}

To see the corrections clearly we determined the effective exponent of 
$\beta$, as the discretized, logarithmic derivative of 
(\ref{Wdef})
\begin{equation}  \label{beff}
\beta_\mathrm{eff}(t) = \frac {\ln W(t,L\to\infty) - \ln W(t',L\to\infty)} 
{\ln(t) - \ln(t^{\prime})} \ .
\end{equation}
using $t/t^{\prime}=2$. As Fig.~\ref{slopes} shows the $\beta_\mathrm{eff}(t)$
curves converge to the same asymptotic value for different sizes 
as $1/t\to 0$, albeit for smaller systems the fluctuations are larger and the 
scaling breaks down as $\xi(t)\simeq L$. One can read-off the most precise
$\beta=0.2415(15)$ estimate from the largest system ($L=2^{17}$), where the 
simulations were followed up to $t=70.000$ MCS. This agrees with our former
estimates for this model \cite{asep2dcikk,asepddcikk}, but now the error 
margin is sufficiently small to exclude a convergence to the field theoretical 
value $\beta=1/4$ \cite{L98} via the analytic corrections (\ref{betacorr}). 

Following the subtraction of the constant leading-order term, corresponding
to $\phi_0=-\beta$, $b_0=1/2$, the remaining corrections are seemingly
small and the oscillations hinder to fit them out very precisely.
We determined the next leading order correction exponent by fitting with
the from (\ref{betaslfit})
\begin{equation}  \label{betaslfit}
\beta_\mathrm{eff}(t) = \beta + b_1\phi_1 t^{\phi_1} \ ,
\end{equation}
in the time window $t>t_\mathrm{mic}$ and before the saturation region. 
From the largest system fit we obtained: $\phi_1=1.05$ and $b_1=-0.12$.

On Fig.~\ref{slopes} we can observe that the local slopes do not change
for late times, therefore assuming a $W\propto t^{0.2415}$ asymptotic
scaling we determined the probability distribution around this law for
the largest size: $L^{17}$ for $1300 < t < 70000$ MCS.
We calculated the distribution of $y = W(t)/t^{0.2415}$ as shown on the
left inset of Fig.~\ref{kpzcoll}.
This opens up the possibility for a future comparison with a solution
like in one dimension \cite{AOS01}.

\begin{figure}
\begin{center}
\epsfxsize=70mm
\epsffile{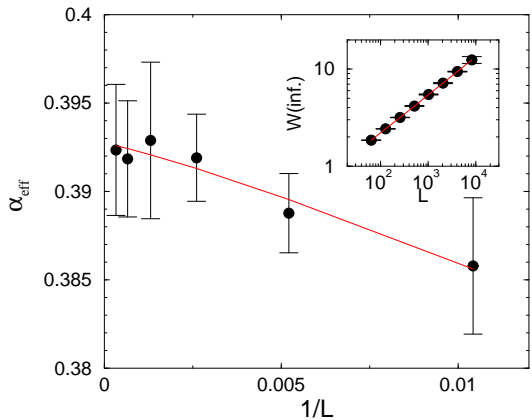}
\caption{(Color online)
Local slopes of the roughness exponent for different sizes. 
The line shows a fit with the form (\ref{WFform}). Inset:
Surface width of the stationary state as the function of linear
system size. 
The line corresponds to a fit with the form (\ref{alphacorr}).}
\label{alphas}
\end{center}
\end{figure}

\begin{figure}
\begin{center}
\epsfxsize=70mm
\epsffile{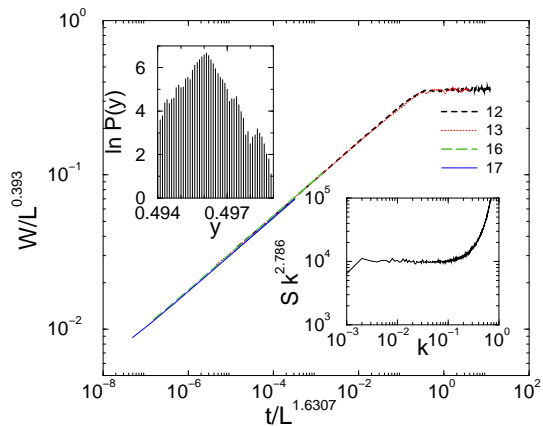}
\caption{(Color online)
Finite size scaling of $W(L,t)$ for $L=2^{12}, 2^{13}, 2^{16}, 2^{17}$.
Right inset: PSD of the $L=2^{13}$ system in the steady state.
Left inset: distribution of $W/t^{0.2415}$ in the growth phase of
the $L=2^{17}$ system.}
\label{kpzcoll}
\end{center}
\end{figure}

Next we investigated the scaling in the steady state. This could be 
achieved in smaller systems with a higher data sampling resolution.
We confirmed that the data corresponds to the steady state by visual
inspection of the $W^2(t,L)$ as well as by analyzing the $P(W^2)$ 
distribution.
Similarly to the time dependence we determined the effective exponent of
the roughness, which can be defined as the logarithmic derivative of
the width
\begin{equation}  \label{aeff}
\alpha_\mathrm{eff}(L) = \frac {\ln W(t\to\infty,L) - \ln W(t\to\infty,L/2)}
{\ln(L) - \ln(L/2)} \ .
\end{equation}
Finite size scaling was done for systems of linear sizes in between
$L_\mathrm{min}=2^{8}$ and $L_\mathrm{max}=2^{13}$ and by
considering the corrections using the fitting form
\begin{equation}  \label{WFform}
\alpha_\mathrm{eff}(L) = \alpha + a_1\omega_1 L^{\omega_1} \ .
\end{equation}
The local slopes of the steady state values $\alpha_\mathrm{eff}(1/L)$ 
are shown on Fig.~\ref{alphas}. The fitting results in: $\alpha=0.393(4)$,
$a_1 = -1.24$ and $\omega_1=1.16$.
Using the $\alpha=0.393(3)$ and $\beta=0.2415(15)$ estimates the dynamical 
scaling exponent is $z=\alpha/\beta=1.627(26)$. With these values we
get $\alpha + z = 2.02(3)$, which satisfies the scaling law expected by 
the Galilean symmetry.
Fig.~\ref{kpzcoll} shows a perfect data collapse with these 
exponents over several decades.

We also investigated the power spectrum density (PSD) of the interface
\begin{equation}\label{PSD}
S({\mathbf k},t) = \langle h({\mathbf k},t) h(-{\mathbf k},t) \rangle  \ ,
\end{equation}
where the height in the Fourier space is computed as
\begin{equation} \label{hF}
h({\mathbf k},t) = \frac{1}{L^{d/2}} \sum_{{\mathbf\chi}} [ h({\mathbf x},t) 
- \langle h\rangle ] \ {\rm exp}(i {\mathbf k \chi}) \ .
\end{equation}
We computed $h({\mathbf k},t)$ from the surface profiles with the FFT method
and determined $S({\mathbf k},t)$ by averaging over $x$ and $y$ directions.
In the steady state the PSD is expected to scale as $S\sim k^{-2-2\alpha}$,
which can be confirmed for $0.002 < k < 0.1$ 
(see inset of Fig.~\ref{kpzcoll}). 
For larger $k$ values we can see a growth of the $S(k)$ function, 
which is the consequence of the lattice regularization.
 
%%%%%%%%%%%%%%%%%%%%%%%%%%%%%%%%%%%%%%%%%%%%%%%%%%%%%%%%%%%%%%%%%%%%%%%%%%
\section{Probability distributions}
%%%%%%%%%%%%%%%%%%%%%%%%%%%%%%%%%%%%%%%%%%%%%%%%%%%%%%%%%%%%%%%%%%%%%%%%%%

The exact form of the spatially averaged height distribution
$P(\langle h\rangle)$ of the KPZ model in one dimension is a hot
topic of statistical physics
\cite{Joh00,PraSp00,Fer05,Sasa04,CD11} and provides a definition
of the KPZ universality class. The $P_L(W^2)$ distribution
in the steady state is also known exactly, in closed
form for small and large $x$ asymptotically \cite{F94}.
In two dimensions not much is known about this distribution.

\begin{figure}
\begin{center}
\epsfxsize=70mm
\epsffile{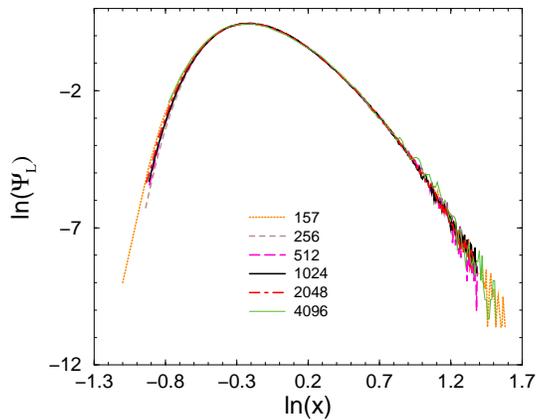}
\caption{(Color online)
The universal scaling function $\Psi_L(x)$ in the steady state
for $L=2^8, 2^9, 2^{10}, 2^{11}, 2^{12}$. 
The dotted line shows the $L=157$ results of \cite{MPPR02}.}
\label{psicomp}
\end{center}
\end{figure}

In \cite{F94,MPPR02} it was shown that the width distributions
$\Psi_L(x) = \langle W^2\rangle P_L(W^2/\langle W^2\rangle)$
of discrete KPZ surfaces exhibit universal behavior.
Now we determined the probability distributions $P_L(W^2)$ and
calculated $\Psi_L(x)$ for $L=2^8, 2^9, 2^{10}, 2^{11}, 2^{12}, 2^{13}$
with the GPU code by measuring $W^2$ in the steady state.
Averaging was done over $N=20-100$ runs and $10^4 - 10^5$ time steps.
In case of the largest, $2^{13}$ case the steady state averaging
was done between $5\times 10^7$ and $10^8$ MCS.
As Fig.~\ref{psicomp} shows the data collapse is very good around
$x=1$, but deviations occur in the large and small $x$
asymptotics due to the lack of sample points there. It is very difficult
to collect enough statistics for the extremal cases as the width 
of $P_L(W^2)$ grows as $L^{2\alpha}$. 

\begin{figure}
\begin{center}
\epsfxsize=70mm
\epsffile{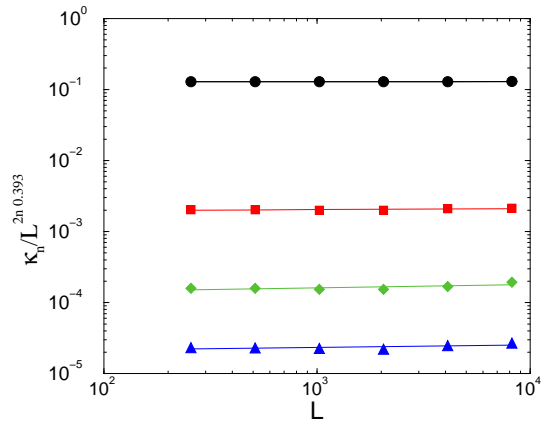}
\caption{(Color online) FSS of the cumulants: 
$\kappa_1$, $\kappa_2$, $\kappa_3$, $\kappa_4$ (top to bottom)
for $L=512, 1024, 2048, 4096, 8192$. The lines correspond to
power-law fitting with very small exponents.}
\label{cum}
\end{center}
\end{figure}

By studying the finite size effects of extreme value statistics
it was discovered \cite{GMOR08} that there is also universality
in the first order (shape) correction to the limit distributions.
It was also shown that if the finite size corrections of the
cumulants can be neglected the shape corrections can be expressed
via the limit distributions.
To see this let us write $P_L(W^2)$ in terms of the cumulant 
generating function
\begin{equation}
P_L(W^2) = \int \frac{dq}{2\pi} \exp{ \Bigl[-iq(W^2-\kappa_1) 
+ \sum_{n=2}^{\infty} \frac{(iq)^n}{n!} \kappa_n \Bigr] } \ ,
\end{equation}
where $\kappa_n$-s are the $n$-th cumulants of $W^2$
(i. e.  $\langle W^2\rangle = \kappa_1 = \nu_1$),
related to the $n$-th, non-central moments ($\nu_n$) as:
\begin{eqnarray}\label{cumulants}
\kappa_1 &=& \nu_1 \ , \nonumber \\
\kappa_2 &=& \nu_2  -  \nu_1^2 \ , \nonumber \\
\kappa_3 &=& \nu_3 - 3 \nu_1 \nu_2 + 2  \nu_1^3 \ , \nonumber \\
\kappa_4 &=& \nu_4 - 4 \nu_3\nu_1 - 3 \nu_2^2 + 12 \nu_1^2\nu_2 - 6 \nu_1^4 \ , \nonumber \\
...
\end{eqnarray} 
Due to general scaling, the cumulants have the large $L$ behavior 
\begin{equation}\label{cumscal}
\kappa_n \propto L^{2 n\alpha}
\end{equation}
Let us assume that the corrections to scaling of the cumulants 
are of the form
\begin{equation}
\kappa_n = L^{2 n\alpha} ( \kappa_n^0 + \kappa_n^1 L^{-\omega_1} + ...)
\end{equation}
To check this we determined the $n=1,2,3,4$ cumulants from 
$W^2$ data and performed a finite size scaling 
analysis. The corrections to scaling (\ref{cumscal}) were 
found to be negligible, as shown on Fig.~\ref{cum}, hence the 
universal limit distribution in principle can very well 
be approximated from the finite $L$ results.

\begin{figure}
\begin{center}
\epsfxsize=70mm
\epsffile{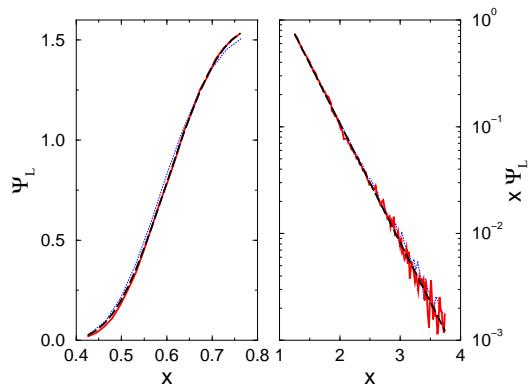}
\caption{(Color online) Small (left) and large (right) asymptotics of
$\Psi_L{(x)}$. Dotted (blue) line corresponds to $L=157$ of 
\cite{F94}, red line $L=2048$ data, dashed line: fitting with the
form (\ref{2dlform}). Right: linear-logarithmic fitting (dashed line)
to the large $x$ part of the same data.}
\label{psi-tail}
\end{center}
\end{figure}

Finally we tried to fit the small and large $x$ asymptotics of 
$\Psi(x)$ with similar functional forms that is known exactly in one 
dimension \cite{F94}.
This assumption is based on the similarity of the underlying model,
i.e. directed migration of dimers instead of particles. 
When we applied for the small $x$ ($x<0.75$) part of 
the $\Psi_L(x)$ the general form 
\begin{equation}\label{1fitform} 
x^A (B-x) \exp{(C/x^D)}
\end{equation}
we found stable nonlinear fittings with $C \simeq 2$ and
$D\simeq 2$ in contrast to one dimension, where $D=1$. 
This is similar to the small $x$ extreme value statistics of 
the $1/f^{\alpha}$ noise, where one obtains $\exp(-a/x^{\beta})$ 
with $\beta$ depending on $\alpha$.
We fixed $C=2$, $D=2$ and tried to get a general form with
integer coefficients up to the second order. We obtained
\begin{equation}\label{2dlform}
x^{-8} (10-x) \exp(-2/x^2) (1+x^{-38} (9-x) \exp(-9.25/x^2))
\end{equation}
in good agreement with the $L=2048$ data as shown on Fig.~\ref{psi-tail}.

For the lager $x$ part we assumed again the form 
of the one dimensional model  
\begin{equation}\label{2drform}
E \exp(F x)/x 
\end{equation}
and obtained a nice agreement with: $E=2.915$, $F=-2.572$.
The least squares fit error was smaller that by a
stretched exponential ansatz.

%%%%%%%%%%%%%%%%%%%%%%%%%%%%%%%%%%%%%%%%%%%%%%%%%%%%%%%%%%%%%%%%%%%%%%%%%%
\section{Conclusions and discussion}
%%%%%%%%%%%%%%%%%%%%%%%%%%%%%%%%%%%%%%%%%%%%%%%%%%%%%%%%%%%%%%%%%%%%%%%%%%

In conclusion we have developed a bit-coded CUDA program, which
simulates very efficiently a $2+1$ dimensional discrete KPZ
growth model (the octahedron model \cite{asep2dcikk}) via binary 
lattice gases.
Using this tool we could achieve extraordinary large sizes up 
to $2^{17} \times 2^{17}$ with a maximum speedup $~240$ on NVIDIA 
Fermi Cards with respect to a single 3 GHz CPU core. 
This allows us to resolve debates over the scaling 
exponents by performing very precise scaling analysis of the 
interface width.

Our growth exponent estimate $\beta=0.2415(15)$ is somewhat bigger
than the results of \cite{Ghai} ($\beta=0.221(2)$), \cite{AA04}
($\beta=0.229(5)$) and \cite{TFW92} ($\beta=0.240(1)$).
It matches our former estimates for this model $0.245(5)$ 
\cite{asepddcikk}, but excludes definitely the $\beta=0.25$ 
field theoretical result. We also estimated the leading
order correction to scaling exponent: $\phi_1=1.05$. 

The independent roughness exponent result $\alpha=0.393(3)$ 
is in the middle of the range obtained by various numerical 
exponent estimates: i.e. between $\alpha=0.36$ \cite{Ghai,Hasel}
$\alpha=0.385(5)$ \cite{TFW92} and the $\alpha=0.4$ field 
theoretical result, well out of error margin. 
This agrees with our former $\alpha=0.395(5)$ \cite{asepddcikk} 
and with the simulation results of \cite{MPP} ($\alpha=0.393(3)$).
In the latter case even the correction to scaling exponent 
$\omega_1=1.16$ is the same.

We analyzed the surface in the steady state by the power spectrum
density method and confirmed the KPZ scaling for several decades above
the lattice cut-off.
We determined the universal scaling function and the cumulants
of the surface width for different sizes and obtained the limit 
distribution via correction to scaling analysis. We gave analytical
fitting for the small and large asymptotics. As compared to 
one dimension \cite{F94}, where a linear $x$ dependence in the
exponential is known exactly, we found $x^2$ dependence 
for small $x$. For the large $x$ deviations the $\exp(F x)/x$ 
tail fits better than a stretched exponential functions
as suggested in \cite{Reis05}.

Our model and code proves to be a very efficient tool to study
not only the $(2+1)$ dimensional KPZ and ASEP models but, more generally
it can be used in the research of fundamental nonequilibrium 
thermodynamical quantities like the large deviation function or
entropy production \cite{DerLDF07,TD10}. It is also straightforward to 
extend it to study more complex system exhibiting pattern formation 
\cite{patscalcikk,iinmproc}, the effect of quenched disorder 
\cite{GPUtexcikk}, the time-dependent structure factor \cite{KS04} 
or to higher dimensions \cite{asepddcikk}.

One may also ask if the results for this discrete model describe those 
of the continuum KPZ equation. In fact this is not an obvious question 
in $d>1$ dimensions \cite{KS04P}. However, we think that the irrelevancy 
of anisotropy by renormalization group studies \cite{TF02} excludes 
this in two dimensions and we find $(2+1)$ dimensional KPZ universality 
class behavior.

On completion of this study we discovered another way of accelerating 
our algorithm, which provides an additional factor of $\sim 1.8$ 
with respect to the simulations presented here. The technical
details will published elsewhere \cite{gpu3cikk}.

\vskip 1.0cm

\noindent
{\bf Acknowledgments:}\\

We thank Zoltan R\'acz and Karl-Heinz Heinig for the useful discussions
and E. Katzav for his comments.
Support from the Hungarian research fund OTKA (Grant No. T077629),
the bilateral German-Hungarian exchange program DAAD-M\"OB 
(Grant Nos. 50450744, P-M\"OB/854) and OSIRIS FP7 is acknowledged. 
The authors thank NVIDIA for supporting the project with high-performance
graphics cards within the framework of Professor Partnership and the
computer center of HZDR for the GPU cluster access.

\bibliography{ws-book9x6}

\end{document}